\documentclass[pra, preprint, floatfix,superscriptaddress]{revtex4}
\usepackage{graphicx}
\usepackage{epstopdf}
\usepackage{amsmath, amsfonts, amssymb, bm}
\begin{document}
\title{Entanglement versus cooling in the system of a driven pair of two-level qubits longitudinally coupled with a boson mode field}
\author{Elena \surname{Cecoi}}
\affiliation{Institute of Applied Physics, Academiei str. 5, MD-2028 Chi\c{s}in\u{a}u, Moldova}
\author{Viorel \surname{Ciornea}}
\affiliation{Institute of Applied Physics, Academiei str. 5, MD-2028 Chi\c{s}in\u{a}u, Moldova}
\author{Aurelian \surname{Isar}}
\affiliation{Horia Hulubei National Institute of Physics and Nuclear Engineering, 
Reactorului str. 30, MG-6 Bucharest - M\u{a}gurele, Romania}
\author{Mihai  A. \surname{Macovei}}
\email{macovei@phys.asm.md}
\affiliation{Institute of Applied Physics, Academiei str. 5, MD-2028 Chi\c{s}in\u{a}u, Moldova}
\date{\today}
\begin{abstract}
The relationship among the entanglement creation within coherently pumped and closely spaced two-level emitters longitudinally coupled with a 
single-mode boson field, and the subsequent quantum cooling of the boson mode is investigated. Even though the two-level qubits are resonantly 
driven, we have demonstrated an efficient cooling mechanism well below limits imposed by the thermal background. Furthermore, the cooling effect 
is accompanied by entanglement of the qubit pair components when the dipole-dipole frequency shift is close to the frequency of the boson mode. 
The maximum boson mode cooling efficiency realizes on the expense of the entanglement creation. Importantly, this occurs for rather weak external 
pumping fields protecting the sample from the deteriorations. Finally, the conditions to effectively optimize these effects are described as well.
\end{abstract}
\maketitle
\section{Introduction}
Entanglement in a few-atom system attracted enormous attention over last few decades \cite{leh1,leh2,ficek,entd1,entd2,lukin,eberly,tq_ent}. 
Small qubit samples may form buildings blocks for even larger networks with huge potential applications for quantum technologies 
\cite{zb_sc,QTBk,agarwal_2014,cirac,chk,wang,kimble,mek_ent,qt_les,qt_pl,qt_m}. Generally, a thorough description of the entanglement 
creation in a two-atom system was given in \cite{fit}. From this point of view artificial atomic systems have been widely investigated as 
well. Particularly, experimental realization of entanglement in two coupled charge qubits was performed in \cite{qcirc}. Entanglement of 
two quantum dots inside a cavity injected with squeezed vacuum was predicted as well, in \cite{gao}. Ultra-strong dipole-dipole 
interacting two-level superconducting flux qubits are naturally entangled through their corresponding environmental reservoir and maximum 
coherence can be induced too \cite{kmj}. Furthermore, a pair of moderately dipole-dipole coupled and laser pumped two-level quantum 
dots get maximal entangled via their environmental phonon thermostat which facilitates also the creation of a subradiant two-qubit state 
\cite{ccim}. Recently, based on quantum dots systems, a relevant experimental realization of an interconnection among two qubits located 
five meters apart from each other, via single photons, was reported in Ref.~\cite{qtsexp}. 

Often to realize quantum states of matter or light one requires ground-state cooled individual or coupled quantum systems, respectively. 
In this context, cooling of a quantum circuit via coupling to an independent or Dicke-like interacting multiqubit ensemble was demonstrated 
in \cite{mmac}. The cooling of a nanomechanical resonator coupled to two interacting flux qubits via the corresponding subradiant Dicke 
states was demonstrated as well, in Ref.~\cite{kxje}. A scheme for ground-state cooling of a mechanical resonator coupled to two 
coupled quantum dots forming an effective $\Lambda$-type three-level structure was  presented in \cite{gxL}. Ground-state cooling of a 
nanomechanical oscillator with $N$ spins was recently proposed in \cite{erem}. 

Evidently, there is a considerable effort done to elucidate the relationship among 
the entanglement and cooling phenomena. For instance, it was found that entanglement enhances cooling in microscopic quantum refrigerators
\cite{popq}. The atom-membrane cooling and entanglement using cavity electromagnetically induced transparency was investigated in 
Ref.~\cite{ritsch}. Ground-state cooling enabled by critical coupling and dark entangled states was found in \cite{gray}. Further interesting 
works on cooling or entanglement processes are presented in Refs.~\cite{c_ex,pt_mar,vitali,beige,popescu,jwang,aurelian}. 

\begin{figure}[t]
\includegraphics[width=8cm]{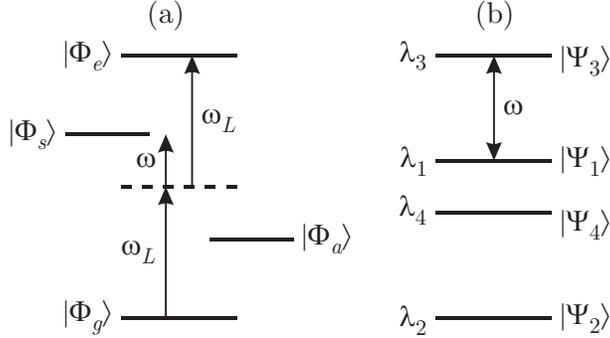}
\caption{\label{fig-0ab} (a) The two-qubit cooperative states $|\Phi_{i}\rangle$, where $i \in \{e,s,a,g\}$, stand for {\it excited}, {\it symmetrical}, 
{\it antisymmetrical} and {\it ground} Dicke states, respectively. A coherent laser source, with its wave-vector perpendicular to the line connecting 
the two qubits, can excite the whole sample either directly, i.e. $|\Phi_{g}\rangle \leftrightarrow |\Phi_{s}\rangle \leftrightarrow |\Phi_{e}\rangle$, 
or indirectly, that is $|\Phi_{g}\rangle \leftrightarrow |\Phi_{e}\rangle$. The latter path involves absorption of two photons simultaneously and is 
less probable for weaker applied external fields - a situation considered here. Therefore, the first channel prevalent meaning that qubit's subsystem 
excitation is taken place via absorption of a single laser photon of frequency $\omega_{L}$ followed, respectively, by a phonon absorption of 
frequency $\omega$. A  process which leads to cooling of the single-mode boson field longitudinally coupled to the two qubits which become also 
entangled. (b) The two-qubit dressed-states $|\Psi_{j}\rangle$ and the corresponding eigenvalues $\lambda_{j}$, $j \in \{1,2,3,4\}$, obtained 
after diagonalization of the dipole-dipole coupled qubit-laser interaction Hamiltonian.}
\end{figure}

Here we shall take the opportunity of these advances and investigate the interconnection of the entanglement process in a coherently and 
resonantly pumped dipole-dipole interacting two-qubit system, and the cooling effects of a boson mode with whom the qubits are longitudinally 
coupled. Both the quantum subsystems are damped via their corresponding environmental reservoirs. We have found that the cooling of the 
boson mode is accompanied by entanglement creation among the two-level qubit pair as long as the dipole-dipole frequency shift lies around 
the boson mode frequency. Generally, the entanglement effect enhances during the cooling process while the amplitude of the applied coherent 
field increases from zero. However, the maximum of the concurrence which was taken as entanglement measure, has slightly lower magnitudes 
than those that would be obtained but in the absence of the boson mode coupling to the two qubits. Thus, the maximum boson mode cooling 
efficiency realizes on the expense of the entanglement creation. Anyway, for lower bath temperatures, the concurrence will increase until 
values which also can be reached without the boson mode coupling to the qubit pair. An intuitive explanation for the described effects is as 
follows: Because we considered that the external coherent source frequency equals the qubit's transition frequency while its wave-vector 
is perpendicular to the line connecting the two dipole-dipole interacting qubits, then a privileged way to excite the collective sample is via a 
simultaneous absorption of a laser photon followed by a boson mode phonon absorption, respectively (see Fig.~\ref{fig-0ab}a). This is 
because the dipole-dipole frequency splitting among the cooperative two-qubit states is close to the boson mode frequency. As a consequence, 
the qubit pair excitation as a whole results in the boson mode cooling, whereas the two closely spaced emitters get entangled. Finally, these 
effects occur for moderately weak external coherent driving fields which can avoid sample's deteriorations.

The article is organized as follows. In Sec. II we describe the system of interest together with the analytical approach used as well as 
the boson mode features. In Sec. III we analyze the entanglement creation among the qubit pair and its relationship with the boson 
mode cooling effects, respectively. We finalize the article with a summary given in Sec. IV.

\section{The master equation and equations of motion}
The system of interest consists of environmental vacuum mediated dipole-dipole coupled pair of identical two-level $\{|2\rangle_{j},|1\rangle_{j}\}$ 
quantum emitters $\{j \in q1,q2\}$, resonantly pumped by a coherent laser source. The laser wavelength is sufficiently bigger than the interqubit 
spatial interval $|\vec r_{q_1q_2}|$, while the interparticle separation is larger than the linear size of the quantum emitter itself. Also, the laser 
wave-vector is perpendicular to the line connecting the qubits. Hence, the qubits are in an equivalent position with respect to the driving field. 
We have assumed that the transition frequencies of both qubits are equal and identical to the external pumping source frequency, respectively. 
Furthermore, the two-level qubits interact with a single boson mode of frequency $\omega$ via a longitudinal coupling. The whole system 
dampens via the interaction with the electromagnetic vacuum modes of the surrounding reservoir as well as through the corresponding boson 
mode environmental thermostat, respectively. As an appropriate realistic model can be taken a coherently pumped two-qubit pair, formed of ions, 
molecules, dimers or impurities as well as quantum dots or quantum wells, superconducting qubits etc., embedded in cavities, nanomechanical resonators 
or quantum circuits ones etc. \cite{blat,les,keel,dimer,zol,sch0,schon,para,twqb_exp,monro,harm,qtsexp,qudits,discord,norri}.

In the following, we shall present the corresponding master equation describing the investigated model where all the involved parameters are included 
properly.
\subsection{The master equation}
The master equation describing this global system in the Born-Markov approximations \cite{leh1,leh2,ficek,zb_sc,agarwal_2014} is given as follows
\begin{eqnarray}
\dot \rho &+&\frac{i}{\hbar}[\bar H,\rho]=-\sum_{\{j,l\} \in\{q1,q2\}}\bigl(\gamma_{jl}[S^{+}_{j},S^{-}_{l}\rho] + H.c.\bigr) \nonumber \\
&-& \frac{\kappa}{2}(1+\bar n)[b^{\dagger},b\rho] - \frac{\kappa}{2}\bar n[b,b^{\dagger}\rho] + H.c.,
\label{ros}
\end{eqnarray}
where an overdot denotes differentiation with respect to time. Here, the qubit operators $S^{+}_{j} = |2\rangle_{j}{}_{j}\langle 1|$, 
$S^{-}_{j}=[S^{+}_{j}]^{\dagger}$ and $S^{(j)}_{z}=(|2\rangle_{j}{}_{j}\langle 2| - |1\rangle_{j}{}_{j}\langle 1|)/2$ obey the commutation relations: 
$[S^{+}_{j},S^{-}_{l}]=2S^{(j)}_{z}\delta_{jl}$ whereas $[S^{(j)}_{z},S^{\pm}_{l}]=\pm S^{\pm}_{j}\delta_{jl}$. The respective boson mode creation, 
$b^{\dagger}$, and annihilation, $b$, operators satisfy the following commutation relations: $[b,b^{\dagger}]=1$ and 
$[b,b]=[b^{\dagger},b^{\dagger}]=0$. In Eq.~(\ref{ros}), the Hamiltonian characterizing the corresponding coherent quantum dynamics is
$\bar H=H + H_{i}$, where
\begin{eqnarray}
H = \hbar\omega b^{\dagger}b + \hbar g\sum_{j\in\{q1,q2\}}S^{(j)}_{z}(b+b^{\dagger}), 
\label{hh}
\end{eqnarray} 
and
\begin{eqnarray}
H_{i} = \hbar\Omega_{dd}\sum_{j \not = l\in \{q1,q2\}}S^{+}_{j}S^{-}_{l} + \hbar\Omega \sum_{j \in \{q1,q2\}}\bigl (S^{+}_{j} + S^{-}_{j}\bigr). 
\label{hi}
\end{eqnarray}
Here, $g$ is the qubit-boson-mode coupling strength whereas $\Omega$ denotes the standard Rabi frequency, both assumed to be identical for each 
qubit, respectively. $\gamma_{q1q1}=\gamma_{q2q2}=\gamma/2$ is the single-qubit spontaneous decay rate, while $\gamma_{q1q2}=\gamma_{q2q1}
=\gamma \chi_{r}/2$ describes the radiative coupling among the two-level qubits and $\Omega_{dd}$ corresponds to the dipole-dipole interaction potential, 
respectively. The radiative coupling $\chi_{r}$ goes to zero (unity) for larger (smaller) interparticle separations $|\vec r_{q_1q_2}|$ in comparison to the 
photon emission wavelength. Correspondingly, $\Omega_{dd}$ tends to zero or to the static dipole-dipole interaction potential. Finally, $\kappa$ is the 
damping rate of the boson mode, while $\bar n$ gives its mean thermal phonon number corresponding to the frequency $\omega$ and environmental 
temperature $T$.

For our further purposes, we diagonalize the Hamiltonian $(\ref{hi})$ describing the dipole-dipole coupled qubit pair interacting as well with an externally 
applied coherent laser field, using the following two-qubit bare states: $|2_{q1}2_{q2}\rangle$, $|2_{q1}1_{q2}\rangle$, $|1_{q1}2_{q2}\rangle$ and 
$|1_{q1}1_{q2}\rangle$. Hence, we arrive at the corresponding cooperative two-qubit eigenfunctions:
\begin{eqnarray}
|\Psi_{4}\rangle &=&-\bar a \{|2_{q1}2_{q2}\rangle + |1_{q1}1_{q2}\rangle\} + \bar b\{|2_{q1}1_{q2}\rangle + |1_{q1}2_{q2}\rangle\}, \nonumber \\
|\Psi_{3}\rangle &=&-\bar c \{|2_{q1}2_{q2}\rangle + |1_{q1}1_{q2}\rangle\} + \bar d\{|2_{q1}1_{q2}\rangle + |1_{q1}2_{q2}\rangle\}, \nonumber \\
|\Psi_{2}\rangle &=&\frac{1}{\sqrt{2}}\{|2_{q1}1_{q2}\rangle - |1_{q1}2_{q2}\rangle\}, \nonumber \\
|\Psi_{1}\rangle &=&\frac{1}{\sqrt{2}}\{|2_{q1}2_{q2}\rangle - |1_{q1}1_{q2}\rangle\}. \label{drst}
\end{eqnarray}
Here 
\begin{eqnarray}
\bar a = \frac{(\Omega_{dd}-\lambda_{4})/\sqrt{2}}{\sqrt{(\Omega_{dd}-\lambda_{4})^{2}+4\Omega^{2}}}, ~~~
\bar b = \sqrt{\frac{2\Omega^{2}}{(\Omega_{dd}-\lambda_{4})^{2}+4\Omega^{2}}}, \nonumber \\
\bar c = \frac{(\Omega_{dd}-\lambda_{3})/\sqrt{2}}{\sqrt{(\Omega_{dd}-\lambda_{3})^{2}+4\Omega^{2}}}, ~~~
\bar d = \sqrt{\frac{2\Omega^{2}}{(\Omega_{dd}-\lambda_{3})^{2}+4\Omega^{2}}}, \nonumber 
\end{eqnarray}
with 
\begin{eqnarray*}
\lambda_{4} &=& \bigl(\Omega_{dd} - \sqrt{\Omega^{2}_{dd}+16\Omega^{2}}\bigr)/2,  \nonumber \\
\lambda_{3} &=& \bigl(\Omega_{dd} + \sqrt{\Omega^{2}_{dd}+16\Omega^{2}}\bigr)/2, 
\end{eqnarray*}
whereas other eigenvalues are 
\begin{eqnarray*}
\lambda_{2} = -\Omega_{dd}, ~~{\rm and}~~ \lambda_{1} = 0, 
\end{eqnarray*}
respectively, see also Fig.~{\ref{fig-0ab}(b)}. Substituting the two-qubit dressed-state transformation (\ref{drst}) in the master equation (\ref{ros}), 
while keeping the slowly varying terms only by assuming that $\omega > g$ with $\omega \approx \lambda_{3}$ as well as $\Omega_{dd} \gg \gamma$ 
and $|\lambda_{4}| \ll |\lambda_{2}|$, one arrives at the following main equation governing the quantum dynamics of the examined system
\begin{eqnarray}
\dot \rho(t) &+& \frac{i}{\hbar}[\bar H_{0},\rho] = -\frac{\gamma}{2}(1+\chi_{r})\biggl(2[\bar c\bar d R_{44} + \bar a\bar b R_{33} + 
\frac{\bar c}{2\sqrt{2}}(R_{41}-R_{14}),\{4(\bar c\bar d R_{44} + \bar a\bar b R_{33}) \nonumber \\
&+& \sqrt{2}\bar c(R_{14} - R_{41})\}\rho] + 2(\bar a \bar d +\bar b \bar c)^{2}\{[R_{34},R_{43}\rho] +[R_{43},R_{34}\rho]\} 
+ \bar a^{2}\{[R_{13},R_{31}\rho] \nonumber \\ 
&+& [R_{31},R_{13}\rho]\} - \sqrt{2}\bar a(\bar a\bar d + \bar b\bar c)\{[R_{43},R_{31}\rho] + [R_{13},R_{34}\rho] - [R_{34},R_{13}\rho] 
- [R_{31},R_{43}\rho]\} \biggr ) \nonumber \\
&-& \frac{\gamma}{2}(1-\chi_{r})\biggl(\bar b^{2}\{[R_{32},R_{23}\rho] + [R_{23},R_{32}\rho]\}+\bar d^{2}\{[R_{24},R_{42}\rho] 
+ [R_{42},R_{24}\rho]\} \nonumber \\
&+& \frac{1}{2}\{[R_{12},R_{21}\rho]+[R_{21},R_{12}\rho]\} - \frac{\bar d}{\sqrt{2}}\{ [R_{42},R_{21}\rho] + [R_{12},R_{24}\rho] 
- [R_{24},R_{12}\rho] \nonumber \\
&-& [R_{21},R_{42}\rho]\} \biggr) - 
\frac{\kappa}{2}(1+\bar n)[b^{\dagger},b\rho] - \frac{\kappa}{2}\bar n[b,b^{\dagger}\rho] + H.c.. \label{dmeq}
\end{eqnarray}
Here
\begin{eqnarray}
\bar H_{0} = \hbar \lambda_{4}R_{44} - \hbar \delta b^{\dagger}b - \hbar \bar g(R_{31}b + b^{\dagger}R_{13}), \label{h0}
\end{eqnarray}
where $\delta=\lambda_{3}-\omega$, whereas $\bar g =\sqrt{2}g\bar c$, see also Fig.~{\ref{fig-0ab}(b)}. The resulting two-qubit dressed-state operators 
which enter in Eq.~(\ref{dmeq}) are defined as follows: $R_{\alpha\beta}=|\Psi_{\alpha}\rangle \langle \Psi_{\beta}|$, $\{\alpha,\beta \in 1, \cdots, 4\}$, 
and satisfy the standard commutation relations $[R_{\alpha\beta},R_{\beta'\alpha'}]=R_{\alpha\alpha'}\delta_{\beta\beta'}$ - 
$R_{\beta'\beta}\delta_{\alpha'\alpha}$. Note that $|\lambda_{4}| \ll |\lambda_{2}|$ means also that we deal with rather weaker applied laser fields, i.e. 
$\Omega/\Omega_{dd} \ll 1$ or the Rabi frequency $\Omega$ is of the order of few $\gamma$'s or even less.

\subsection{The equations of motion}
Using the Master equation (\ref{dmeq}), one can obtain the following equations of motion describing the combined {\it laser pumped qubit pair plus boson 
mode} sample where the corresponding pumping and damping effects are properly taken into account:
\begin{eqnarray}
\dot P^{(0)}_{n} &=& i\bar g(P^{(4)}_{n} - P^{(6)}_{n}) - \kappa\bar n\bigl((n+1)P^{(0)}_{n} - nP^{(0)}_{n-1}\bigr) \nonumber \\
&-& \kappa(1+\bar n)\bigl(nP^{(0)}_{n} - (n+1)P^{(0)}_{n+1}\bigr), \nonumber \\
\dot P^{(1)}_{n} &=& i\bar g P^{(4)}_{n} - \kappa\bar n\bigl((n+1)P^{(1)}_{n} - nP^{(1)}_{n-1}\bigr) \nonumber \\
&-&\kappa(1+\bar n)\bigl(nP^{(1)}_{n} - (n+1)P^{(1)}_{n+1}\bigr) + \gamma^{(1)}_{0}P^{(0)}_{n} \nonumber \\
&-& \gamma^{(1)}_{1}P^{(1)}_{n} - \gamma^{(1)}_{2}P^{(2)}_{n} - \gamma^{(1)}_{3}P^{(3)}_{n} +
\gamma^{(1)}_{11}P^{(11)}_{n}, \nonumber 
\end{eqnarray}
\begin{eqnarray}
\dot P^{(2)}_{n} &=& - \kappa(1+\bar n)\bigl(nP^{(2)}_{n} - (n+1)P^{(2)}_{n+1}\bigr) \nonumber \\
&-&  \kappa\bar n\bigl((n+1)P^{(2)}_{n}- nP^{(2)}_{n-1}\bigr) + \gamma^{(2)}_{0}P^{(0)}_{n} \nonumber \\
&+& \gamma^{(2)}_{1}P^{(1)}_{n} - \gamma^{(2)}_{2}P^{(2)}_{n} + \gamma^{(2)}_{3}P^{(3)}_{n} - \gamma^{(2)}_{11}P^{(11)}_{n}, \nonumber \\
\dot P^{(3)}_{n} &=& -i\bar g P^{(6)}_{n} + \gamma^{(3)}_{0}P^{(0)}_{n} + \gamma^{(3)}_{1}P^{(1)}_{n} -  \gamma^{(3)}_{2}P^{(2)}_{n}  \nonumber  \\
&-& \gamma^{(3)}_{3}P^{(3)}_{n} - \gamma^{(3)}_{11}P^{(11)}_{n} -\kappa\bar n\bigl((n+1)P^{(3)}_{n}  \nonumber \\
&-& nP^{(3)}_{n-1}\bigr) - \kappa(1+\bar n)\bigl(nP^{(3)}_{n}-(n+1)P^{(3)}_{n+1}\bigr), \nonumber \\
\dot P^{(4)}_{n} &=& -i\delta P^{(5)}_{n} +2 i\bar g n (P^{(1)}_{n} - P^{(3)}_{n-1} ) - \kappa(1+\bar n)\bigl(2P^{(6)}_{n} \nonumber \\
&+&(2n-1)P^{(4)}_{n} -2 (n+1)P^{(4)}_{n+1}\bigr)/2 + \kappa\bar n\bigl(2nP^{(4)}_{n-1} \nonumber \\
&-& (2n+1)P^{(4)}_{n} \bigr)/2 - \gamma^{(4)}_{4}P^{(4)}_{n} + \gamma^{(4)}_{8}P^{(8)}_{n}, \nonumber \\
\dot P^{(5)}_{n} &=& -i\delta P^{(4)}_{n}  - \kappa(1+\bar n)\bigl(2P^{(7)}_{n} + (2n-1)P^{(5)}_{n} \nonumber \\
&-&2 (n+1)P^{(5)}_{n+1}\bigr)/2 - \kappa\bar n\bigl( (2n+1)P^{(5)}_{n} \nonumber \\
&-& 2nP^{(5)}_{n-1} \bigr)/2 - \gamma^{(5)}_{5}P^{(5)}_{n} + \gamma^{(5)}_{9}P^{(9)}_{n}, \nonumber \\
\dot P^{(6)}_{n} &=& -i\delta P^{(7)}_{n} +2 i\bar g(n+1)(P^{(1)}_{n+1} - P^{(3)}_{n}) - \kappa(1+\bar n) \nonumber \\
&\times& \bigl((2n+1)P^{(6)}_{n} -2 (n+1)P^{(6)}_{n+1}\bigr)/2 + \kappa\bar n\bigl(2nP^{(6)}_{n-1} \nonumber \\
&-& (2n+3)P^{(6)}_{n} + 2P^{(4)}_{n}\bigr)/2 - \gamma^{(6)}_{6}P^{(6)}_{n} + \gamma^{(6)}_{12}P^{(12)}_{n}, \nonumber \\
\dot P^{(7)}_{n} &=& -i\delta P^{(6)}_{n} - \kappa(1+\bar n)\bigl((2n+1)P^{(7)}_{n} -2 (n+1) \nonumber \\
&\times&P^{(7)}_{n+1}\bigr)/2 -\kappa\bar n\bigl( (2n+3)P^{(7)}_{n}- 2nP^{(7)}_{n-1}  \nonumber \\
&-&2P^{(5)}_{n} \bigr)/2  -\gamma^{(7)}_{7}P^{(7)}_{n} + \gamma^{(7)}_{13}P^{(13)}_{n}, \nonumber \\
\dot P^{(8)}_{n} &=&i(\lambda_{4}-\delta)P^{(9)}_{n} + i\bar g n P^{(11)}_{n} - \kappa\bar n\bigl( (2n+1)P^{(8)}_{n} \nonumber \\
&-&2nP^{(8)}_{n-1} \bigr)/2 - \kappa(1+\bar n)\bigl( (2n-1)P^{(8)}_{n} + 2P^{(12)}_{n}\nonumber \\
&-&2 (n+1)P^{(8)}_{n+1}\bigr)/2 + \gamma^{(8)}_{4}P^{(4)}_{n} - \gamma^{(8)}_{8}P^{(8)}_{n}, \nonumber \\
\dot P^{(9)}_{n} &=&i(\lambda_{4}-\delta)P^{(8)}_{n} + i\bar g n P^{(10)}_{n} - \kappa\bar n\bigl( (2n+1)P^{(9)}_{n} \nonumber \\
&-&2nP^{(9)}_{n-1} \bigr)/2 - \kappa(1+\bar n)\bigl((2n-1)P^{(9)}_{n}+2P^{(13)}_{n}\nonumber \\
&-&2(n+1)P^{(9)}_{n+1}\bigr)/2 + \gamma^{(9)}_{5}P^{(5)}_{n} - \gamma^{(9)}_{9}P^{(9)}_{n}, \nonumber \\
\dot P^{(10)}_{n} &=&i\lambda_{4}P^{(11)}_{n}+ i\bar g P^{(9)}_{n} - \kappa\bar n\bigl((n+1)P^{(10)}_{n} - nP^{(10)}_{n-1}\bigr) \nonumber \\
&-&\kappa(1+\bar n)\bigl(nP^{(10)}_{n} - (n+1)P^{(10)}_{n+1}\bigr) - \gamma^{(10)}_{10}P^{(10)}_{n}, \nonumber \\
\dot P^{(11)}_{n} &=&i\lambda_{4}P^{(10)}_{n}+ i\bar g P^{(8)}_{n} - \kappa\bar n\bigl((n+1)P^{(11)}_{n} - nP^{(11)}_{n-1}\bigr) \nonumber \\
&-&\kappa(1+\bar n)\bigl(nP^{(11)}_{n} - (n+1)P^{(11)}_{n+1}\bigr) + \gamma^{(11)}_{0}P^{(0)}_{n} \nonumber 
\end{eqnarray}
\begin{eqnarray}
&-& \gamma^{(11)}_{1}P^{(1)}_{n} - \gamma^{(11)}_{2}P^{(2)}_{n} - \gamma^{(11)}_{3}P^{(3)}_{n} - \gamma^{(11)}_{11}P^{(11)}_{n}, \nonumber \\
\dot P^{(12)}_{n}&=& i(\lambda_{4}-\delta)P^{(13)}_{n} + i\bar g(n+1) P^{(11)}_{n+1} + \gamma^{(12)}_{6}P^{(6)}_{n} \nonumber \\
&-& \kappa(1+\bar n)\bigl((2n+1)P^{(12)}_{n}- 2(n+1)P^{(12)}_{n+1} \bigr)/2  \nonumber \\
&-& \kappa\bar n\bigl((2n+3) P^{(12)}_{n} -2nP^{(12)}_{n-1} - 2P^{(8)}_{n}\bigr)/2  \nonumber \\
&-& \gamma^{(12)}_{12}P^{(12)}_{n}, \nonumber \\
\dot P^{(13)}_{n}&=& i(\lambda_{4}-\delta)P^{(12)}_{n} + i\bar g(n+1) P^{(10)}_{n+1} + \gamma^{(13)}_{7}P^{(7)}_{n} \nonumber \\
&-& \kappa(1+\bar n)\bigl((2n+1)P^{(13)}_{n}- 2(n+1)P^{(13)}_{n+1} \bigr)/2  \nonumber \\
&-& \kappa\bar n\bigl((2n+3) P^{(13)}_{n} -2nP^{(13)}_{n-1} - 2P^{(9)}_{n}\bigr)/2  \nonumber \\
&-& \gamma^{(13)}_{13}P^{(13)}_{n}. \label{eqsm}
\end{eqnarray}
The system of equations (\ref{eqsm}) can be easily obtained if one first get the equations of motion for the variables  
$\rho_{\alpha\beta}=\langle \alpha|\rho|\beta\rangle$, $\{\alpha,\beta \in 1, \cdots, 4\}$, see also \cite{quang}, using the Master Equation (\ref{dmeq}), 
namely, $\rho^{(0)}=\rho_{11}+\rho_{22}+\rho_{33}+\rho_{44}$, $\rho^{(1)}=\rho_{11}$, $\rho^{(2)}=\rho_{22}$, 
$\rho^{(3)}=\rho_{33}$, $\rho^{(4)}=b^{\dagger}\rho_{31}-\rho_{13}b$, $\rho^{(5)}=b^{\dagger}\rho_{31} +\rho_{13}b$, 
$\rho^{(6)}=\rho_{31}b^{\dagger} - b\rho_{13}$, $\rho^{(7)}=\rho_{31}b^{\dagger} + b\rho_{13}$, 
$\rho^{(8)}=b^{\dagger}\rho_{34}-\rho_{43}b$, $\rho^{(9)}=b^{\dagger}\rho_{34} + \rho_{43}b$,
$\rho^{(10)}=\rho_{14}-\rho_{41}$, $\rho^{(11)}=\rho_{14}+\rho_{41}$, $\rho^{(12)}=\rho_{34}b^{\dagger} - b\rho_{43}$,  
and $\rho^{(13)}=\rho_{34}b^{\dagger} + b\rho_{43}$. The projection on the Fock states $|n\rangle$, i.e., 
$P^{(i)}_{n}=\langle n|\rho^{(i)}|n\rangle$, $\{i \in 0, \cdots, 13\}$, with $n \in \{0, \infty\}$, will lead us to Eqs.~(\ref{eqsm}). 
The corresponding decay rates are given in the Appendix A. 
\begin{figure}[t]
\includegraphics[width=4.56cm]{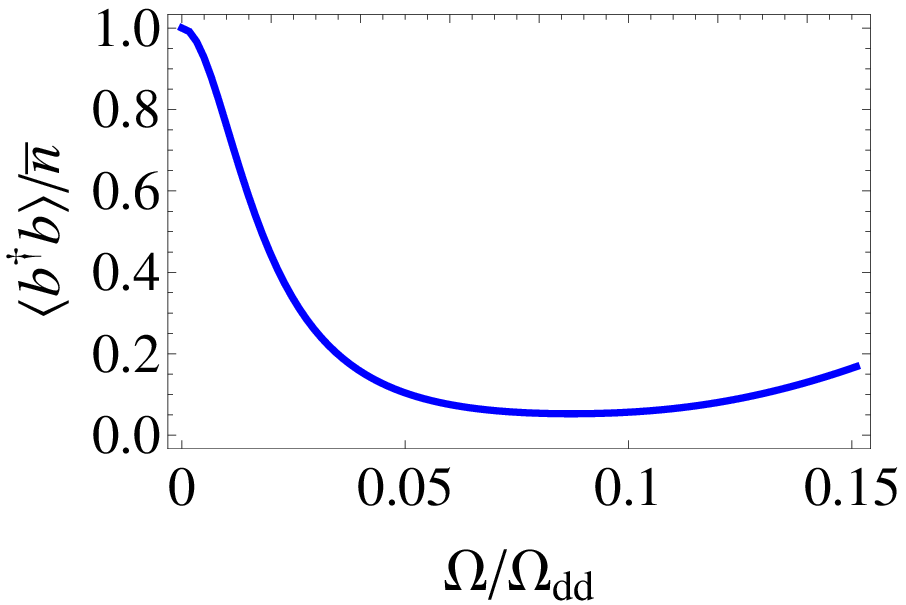}
\includegraphics[width=4.58cm]{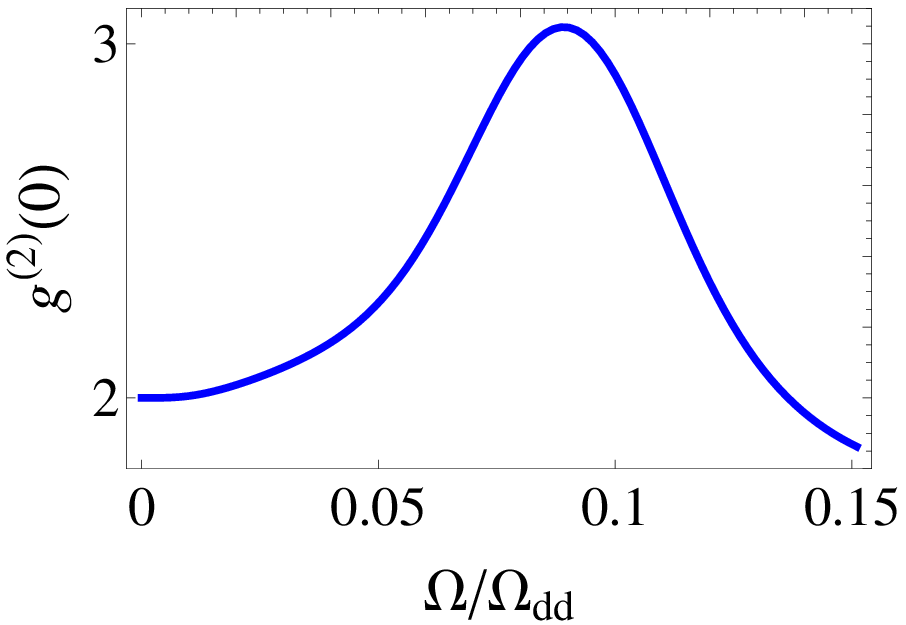}
\begin{picture}(0,0)
\put(-160,70){(a)}
\put(-25,70){(b)}
\end{picture}
\caption{\label{fig-1ab} (a) The steady-state behavior of the scaled mean phonon number $\langle b^{\dagger}b\rangle/\bar n$ 
as well as (b) the steady-state behavior of the second-order phonon-phonon correlation function $g^{(2)}(0)$ versus $\Omega/\Omega_{dd}$, 
respectively. The involved parameters are: $g/\gamma=2$, $\Omega_{dd}/\gamma=28$, $\omega/\gamma=30$, $\chi_{r}=0.98$, 
$\bar n=20$ and $\kappa/\gamma=10^{-3}$.}
\end{figure}

Generally, in order to solve the infinite system of equations (\ref{eqsm}), one truncates it at a certain maximum value $n=n_{max}$ so that a further 
increase of its value, i.e. $n_{max}$, does not modify the obtained results. As a consequence, the steady-state mean phonon number is expressed as:
\begin{eqnarray}
\langle b^{\dagger} b \rangle = \sum^{n_{max}}_{n=0}nP_{n}^{(0)}, \label{nfm}
\end{eqnarray}
with
\begin{eqnarray}
\sum^{n_{max}}_{n=0}P_{n}^{(0)}=1, \label{nrm}
\end{eqnarray}
while its steady-state second-order phonon-phonon correlation function is defined in the usual way \cite{glauber}, namely,
\begin{eqnarray}
g^{(2)}_{b}(0) &=& \frac{ \langle b^{\dagger}b^{\dagger}bb\rangle }{\langle b^{\dagger} b\rangle^{2} } \nonumber \\
&=&\frac{1}{\langle b^{\dagger}b\rangle ^{2}}\sum^{n_{max}}_{n=0}n(n-1)P_{n}^{(0)}. \label{gg2}
\end{eqnarray}

\begin{figure}[t]
\includegraphics[width=4.56cm]{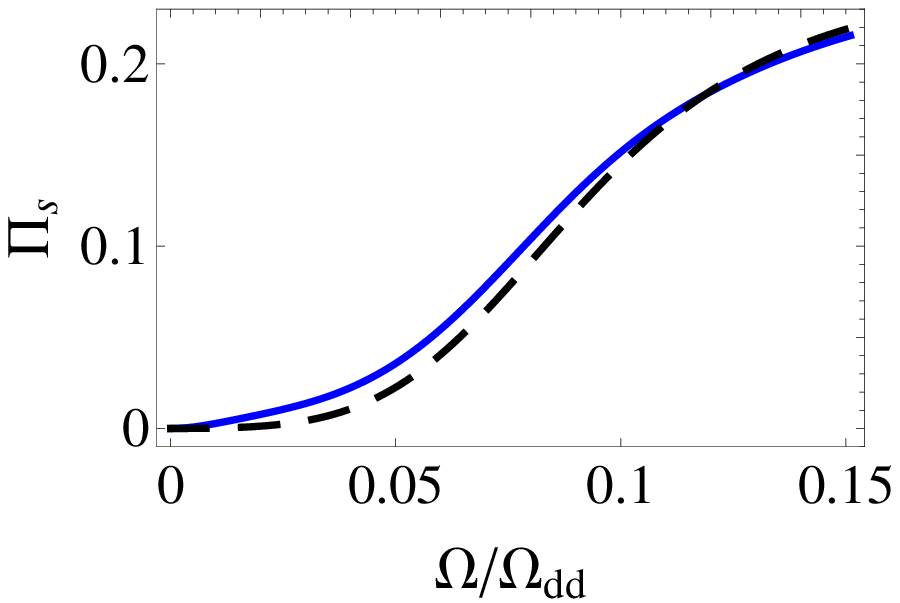}
\includegraphics[width=4.58cm]{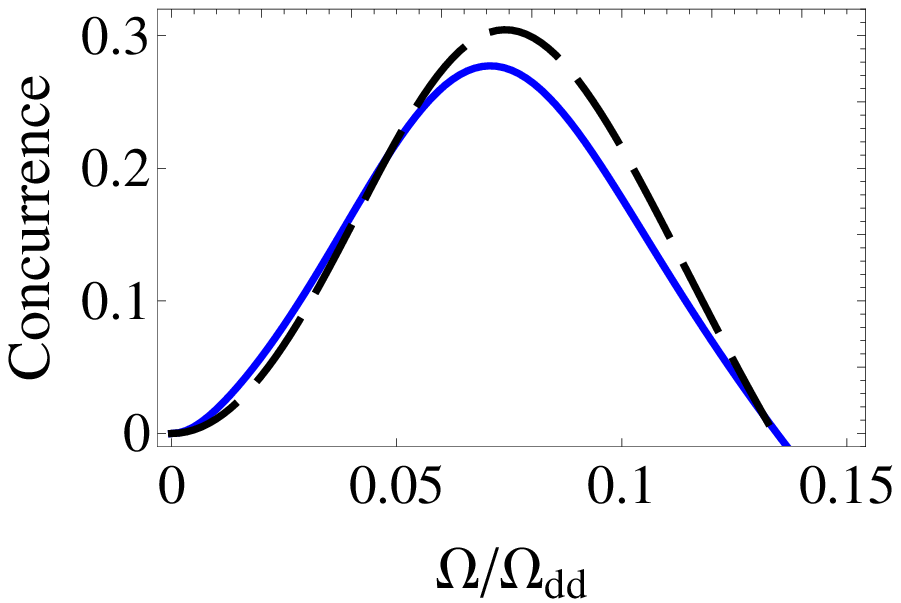}
\begin{picture}(0,0)
\put(-240,70){(a)}
\put(-25,70){(b)}
\end{picture}
\caption{\label{fig-2ab} (a) The steady-state behaviors of the population in the symmetrical two-qubit Dicke state 
$|\Phi_{s}\rangle$=$\{|2_{q1}1_{q2}\rangle + |1_{q1}2_{q2}\rangle\}/\sqrt{2}$, i.e. 
$\Pi_{s} =\langle|\Phi_{s}\rangle \langle \Phi_{s}|\rangle$, as a function of scaled pumping parameter $\Omega/\Omega_{dd}$. 
(b) The corresponding behaviors but for the two-qubit concurrence $C$. In these plots the solid lines are plotted for $g\not =0$
while the dashed curves describe the same results but with $g=0$, respectively. All other parameters are the same as for 
Fig.~(\ref{fig-1ab}).}
\end{figure}

Based on Eqs.~(\ref{eqsm},\ref{nfm},\ref{nrm},\ref{gg2}), Figure \ref{fig-1ab}(a) shows the cooling of the boson mode while the pumping parameter 
is being varying demonstrating an efficient cooling scheme. Respectively, Figure \ref{fig-1ab}(b) depicts the second-order phonon-phonon 
correlation function, during the cooling process, demonstrating super-Poissonian phonon statistics, i.e. $g^{(2)}(0) > 2$, with only few phonons. 
The cooling mechanism occurring in this system, when $\Omega \ll \Omega_{dd}$, can be intuitively understood if one refers to the two-qubit Dicke 
states \cite{leh1,leh2,ficek}, namely, $|\Phi_{e}\rangle=|2_{q1}2_{q2}\rangle$, $|\Phi_{s}\rangle$=$\{|2_{q1}1_{q2}\rangle + 
|1_{q1}2_{q2}\rangle\}/\sqrt{2}$, $|\Phi_{a}\rangle$=$\{|2_{q1}1_{q2}\rangle - |1_{q1}2_{q2}\rangle\}/\sqrt{2}$, and $|\Phi_{g}\rangle=
|1_{q1}1_{q2}\rangle$, see Fig.~\ref{fig-0ab}(a). When the external field frequency is in resonance with that of the qubit one, while its 
wave-vector is perpendicular to the line connecting the two qubits, then the only way to laser excite the two-qubit sample is either via 
$|\Phi_{g}\rangle \to |\Phi_{s}\rangle \to |\Phi_{e}\rangle$ or $|\Phi_{g}\rangle \to |\Phi_{e}\rangle$, respectively. The latter path involve 
two-photon processes which are less probable for weaker external driving fields. Therefore, the first channel, i.e. 
$|\Phi_{g}\rangle \to |\Phi_{s}\rangle \to |\Phi_{e}\rangle$, becomes active and involves an available phonon at a particular frequency from 
simple reasons since the driving external coherent field is in resonance with the qubit's transition frequency, that is $\omega_{sg} \approx \omega_{L} 
+ \omega$, see Fig.~\ref{fig-0ab}(a). Thus, one can conclude that the two-qubit system absorbs an external laser photon followed by a boson mode 
phonon absorption in order to reach the symmetrical Dicke state $|\Phi_{s}\rangle$, and these processes lead to phonon cooling effects, respectively. 
Figure \ref{fig-2ab}(a) demonstrates this statement in the sense that the population of the symmetrical two-qubit state, i.e. $|\Phi_{s}\rangle$, 
increases in the presence of the boson mode, coupled to the two qubits, while compared to the case of its absence. In the dressed-state picture, 
depicted in Fig.~\ref{fig-0ab}(b), cooling occurs evidently when the population in the dressed-state $|\Psi_{1}\rangle$ is larger than that residing 
in the state $|\Psi_{3}\rangle$, respectively. If one inspects the explicit forms of these states, see Exps.~(\ref{drst}), then one can observe that 
laser excitation involves directly all the two-qubit Dicke states, excepting the antisymmetrical one, i.e. $|\Phi_{a}\rangle$.

Until now we have focused mainly on the boson mode properties. In the following Section using the analytical approach developed here, we shall 
investigate the entanglement creation within the qubit subsystem and emphasize its connection to the boson mode cooling phenomena, respectively. 
\section{Entanglement of the two-qubit system coupled with a single-mode boson field}
The entanglement and its definition is certainly a main topic within quantum computation theory. In this context, one of the widely accepted measures 
of entanglement for a two qubit system is the concurrence, $C$, \cite{entd1,entd2}. Particularly, for a mixed state of qubits $\{q_{1},q_{2}\}$ with 
density matrix $\tilde \rho_{q_{1}q_{2}}$, it is defined as 
\begin{equation}
C=\rm{max}\{0,s_{1}-\sum^{4}_{\xi=2}s_{\xi}\}. \label{ccr}
\end{equation}
The quantities $s_{\xi}$, $\{ \xi \in 1,\cdots,4\}$, are the square roots of the eigenvalues of the following matrix product
\begin{equation}
Q=\tilde \rho_{q_{1}q_{2}}(\sigma_{q_{1}y}\otimes\sigma_{q_{2}y})\tilde \rho^{\ast}_{q_{1}q_{2}}(\sigma_{q_{1}y}\otimes\sigma_{q_{2}y}),
\label{mp}
\end{equation}
and, importantly, in descending order. Here, $\tilde \rho^{\ast}_{q_{1}q_{2}}$ denotes complex conjugation of $\tilde \rho_{q_{1}q_{2}}$, and 
$\sigma_{jy}$ are Pauli matrices for the two-level systems ($j \in \{q_{1},q_{2}\}$). The values of the concurrence range from zero for an unentangled 
state to unity for a maximally entangled two-particle state \cite{entd1,entd2}. The density matrix $\tilde \rho_{q_1q_2}$ can be represented in the basis 
$|2_{q1}2_{q2}\rangle$, $|2_{q1}1_{q2}\rangle$, $|1_{q1}2_{q2}\rangle$ and $|1_{q1}1_{q2}\rangle$, which is symmetric under the exchange of the 
sub-systems \cite{entd1,entd2,wang,mek_ent}. Hence, its elements are given as follows:
\begin{eqnarray}
\tilde \rho_{q_1q_2} = \left( \begin{array}{cccc}
\tilde \rho_{11} & \tilde \rho_{12} & \tilde \rho_{13} & \tilde \rho_{14}\\
\tilde \rho_{21} & \tilde \rho_{22} & \tilde \rho_{23} & \tilde \rho_{24}\\
\tilde \rho_{31} & \tilde \rho_{32} & \tilde \rho_{33} & \tilde \rho_{34} \\ 
\tilde \rho_{41} & \tilde \rho_{42} & \tilde \rho_{43} & \tilde \rho_{44}
\end{array} \right), 
\label{rab}
\end{eqnarray}
where
\begin{eqnarray}
\tilde \rho_{11} &=& \frac{1}{4}(1+\rho_{11}-\rho_{22}) - \frac{\Omega_{dd}}{4\sqrt{\Omega^{2}_{dd} +(4\Omega)^{2}}}(\rho_{33}-\rho_{44}) 
\nonumber \\
&-&\frac{\bar a}{\sqrt{2}}(\rho_{41}+\rho_{14}), \nonumber \\
\tilde \rho_{12} &=&\frac{\Omega}{\sqrt{\Omega^{2}_{dd} +(4\Omega)^{2}}}(\rho_{33}-\rho_{44}) - \frac{\bar c}{\sqrt{2}}\rho_{41}, \nonumber \\
\tilde \rho_{13} &=&\tilde \rho_{12}, \nonumber \\
\tilde \rho_{14} &=&\frac{1}{4}(1+\frac{\Omega_{dd}}{\sqrt{\Omega^{2}_{dd} +(4\Omega)^{2}}})\rho_{44} - \frac{\bar a}{\sqrt{2}}(\rho_{41}-\rho_{14})
\nonumber \\
&+& \frac{1}{4}(1-\frac{\Omega_{dd}}{\sqrt{\Omega^{2}_{dd} +(4\Omega)^{2}}})\rho_{33} - \frac{1}{2}\rho_{11}, \nonumber \\
\tilde \rho_{21} &=&(\tilde \rho_{12})^{\dagger}, \nonumber \\
\tilde \rho_{22} &=& \frac{1}{4}(1+\rho_{22}-\rho_{11}) + \frac{\Omega_{dd}}{4\sqrt{\Omega^{2}_{dd} +(4\Omega)^{2}}}(\rho_{33}-\rho_{44}), 
\nonumber \\
\tilde \rho_{23} &=&\frac{1}{4}(1-\frac{\Omega_{dd}}{\sqrt{\Omega^{2}_{dd} +(4\Omega)^{2}}})\rho_{44} - \frac{1}{2}\rho_{22}
\nonumber \\
&+& \frac{1}{4}(1 + \frac{\Omega_{dd}}{\sqrt{\Omega^{2}_{dd}+(4\Omega)^{2}}})\rho_{33}, \nonumber \\
\tilde \rho_{24} &=&\frac{\Omega}{\sqrt{\Omega^{2}_{dd} +(4\Omega)^{2}}}(\rho_{33}-\rho_{44}) + \frac{\bar c}{\sqrt{2}}\rho_{14}, \nonumber \\
\tilde \rho_{31} &=&(\tilde \rho_{13})^{\dagger}, ~~~ \tilde \rho_{32} = \tilde \rho_{23}, ~~~
\tilde \rho_{33} = \tilde \rho_{22}, ~~~ \tilde \rho_{34} =\tilde \rho_{24}, \nonumber \\
\tilde \rho_{41} &=&(\tilde \rho_{14})^{\dagger}, ~~~ \tilde \rho_{42} =(\tilde \rho_{24})^{\dagger}, ~~~ \tilde \rho_{43}=(\tilde \rho_{34})^{\dagger},
\nonumber 
\end{eqnarray}
\begin{eqnarray}
\tilde \rho_{44} &=& \frac{1}{4}(1+\rho_{11}-\rho_{22}) - \frac{\Omega_{dd}}{4\sqrt{\Omega^{2}_{dd} +(4\Omega)^{2}}}(\rho_{33}-\rho_{44}) 
\nonumber \\
&+&\frac{\bar a}{\sqrt{2}}(\rho_{41}+\rho_{14}). \label{roq12}
\end{eqnarray}
Inserting the matrix (\ref{rab}) in the expression (\ref{mp}) one can obtain the corresponding eigenvalues of $Q$ after some algebraic manipulations. 
The corresponding steady-state behaviors for the concurrence $C$ are shown in the Figure~\ref{fig-2ab}(b) which were obtained with the help of 
Eqs.~(\ref{eqsm}) as well as Exps.~(\ref{ccr}-\ref{roq12}). One can observe that the entanglement creation among the two-level qubit pair is 
accompanied by cooling of the boson mode (compare the solid curves in Fig.~\ref{fig-1ab}(a) and Fig.~\ref{fig-2ab}(b), respectively). Moreover, 
the maximal cooling effect is achieved when the entanglement is maximal as well. Note that even higher magnitudes for the concurrence $C$ can be 
obtained but in the absence of the qubit's coupling to the single-mode boson field (compare the solid and dashed curves in Fig.~\ref{fig-2ab}b) meaning 
that the maximal cooling efficiency realizes on the expense of the entanglement creation. Anyway, at the beginning of the steady-state evolution the 
concurrence $C$ is larger than its value which would be obtained, however, in the absence of the boson mode, see Fig.~\ref{fig-2ab}(b). Furthermore, 
for lower bath temperatures, the magnitude of the concurrence $C$ will reach the same values regardless of the boson mode presence. This is because 
at those temperatures the corresponding two-qubit cooperative states, responsible for entanglement creation, are almost equal populated in both cases.

\section{Summary}
Summarizing, we have investigated the relationship among the entanglement creation in a laser-pumped dipole-dipole interacting two-level qubits 
and the cooling effects of a boson mode which is longitudinally coupled with the both quantum emitters, respectively. We have found that cooling 
occurs when the dipole-dipole frequency shift lies around the boson mode frequency. This happens because it was assumed that the driving coherent 
field frequency is equal with the qubit's transition one, while the two qubits are in an equivalent position with respect to the pumping external field.
Hence, the only way to excite the two-qubit sample is via a concomitant absorption of a photon and a phonon, respectively, leading to cooling of the 
boson mode. Furthermore, the quantum cooling process is accompanied by entanglement creation within the qubit sample which is demonstrated by 
nonzero values for the concurrence, although its maximal magnitude, i.e. for $C$, is lower than that which would be obtained but in the absence of 
the single-mode boson field. However, adjusting the external parameters one can optimize the entanglement as well. These effects are taken place 
for rather weak external applied fields which may protect the sample from deteriorations.

This work was supported by grant No. 15.817.02.09F. Also, M.M.A. is grateful for the nice hospitality of the Theory Department of the 
Horia Hulubei National Institute of Physics and Nuclear Engineering, Bucharest, Romania. 

\appendix
\section{The decay rates entering in the equations of motion (\ref{eqsm})}
Below one can find the corresponding decay rates which enter in the Eqs.~(\ref{eqsm}), that is,
$\gamma^{(1)}_{0}=\gamma\bar c^{2}(1+\chi_{r})$, $\gamma^{(1)}_{1}=\gamma\{(\bar a^{2}+2\bar c^{2})(1+\chi_{r}) + (1-\chi_{r})/2\}$,
$\gamma^{(1)}_{2}=\gamma\{\bar c^{2}(1+\chi_{r}) - (1-\chi_{r})/2\}$, $\gamma^{(1)}_{3}=\gamma(1+\chi_{r})(\bar c^{2} - \bar a^{2})$,
$\gamma^{(1)}_{11}=\gamma\{(1+\chi_{r})(\bar a(\bar a\bar d + \bar b \bar c)/\sqrt{2} + \sqrt{2}\bar d\bar c^{2}) + \frac{\bar d}{2\sqrt{2}}(1-\chi_{r})\}$,
$\gamma^{(2)}_{0}=\gamma\bar d^{2}(1-\chi_{r})$, $\gamma^{(2)}_{1}=\gamma(1-\chi_{r})(1/2-\bar d^{2})$, 
$\gamma^{(2)}_{2}=\gamma(1-\chi_{r})(1/2 + \bar b^{2}+2\bar d^{2})$, $\gamma^{(2)}_{3}=\gamma(1-\chi_{r})(\bar b^{2} -\bar d^{2})$, 
$\gamma^{(2)}_{11}=\gamma \bar d(1-\chi_{r})/\sqrt{2}$, $\gamma^{(3)}_{0}=2\gamma(\bar a \bar d + \bar b \bar c)^{2}(1 + \chi_{r})$,
$\gamma^{(3)}_{1}=\gamma\{\bar a^{2} - 2(\bar a\bar d+\bar b\bar c)^{2}\}(1 + \chi_{r})$, $\gamma^{(3)}_{2}=\gamma\{2(\bar a\bar d+\bar b\bar c)^{2}
(1 + \chi_{r}) - \bar b^{2}(1-\chi_{r})\}$, $\gamma^{(3)}_{3}=2\gamma\{(2(\bar a\bar d+\bar b\bar c)^{2} + \bar a^{2}/2)(1 + \chi_{r}) 
+ \bar b^{2}(1-\chi_{r})/2\}$, $\gamma^{(3)}_{11}=\sqrt{2}\bar a\gamma(\bar a\bar d + \bar b\bar c)(1+\chi_{r})$, 
$\gamma^{(4)}_{4}=\gamma\{(4(\bar a\bar b)^{2} + (\bar a\bar d + \bar b\bar c)^{2} + \bar a^{2} + \bar c^{2}/2)(1+\chi_{r})+(1/2+\bar b^{2})(1-\chi_{r})/2\}$,
$\gamma^{(4)}_{8}=\gamma\{(\sqrt{2}\bar c(2\bar a\bar b + \bar c\bar d)+\bar a(\bar a\bar d+ \bar b\bar c)/\sqrt{2})(1+\chi_{r})+\frac{\bar d}{2\sqrt{2}}(1-\chi_{r})\}$,
$\gamma^{(5)}_{5}=\gamma^{(4)}_{4}$, $\gamma^{(5)}_{9}=\gamma^{(4)}_{8}$, $\gamma^{(6)}_{6}=\gamma^{(5)}_{5}$, $\gamma^{(6)}_{12}=\gamma^{(5)}_{9}$,
$\gamma^{(7)}_{7}=\gamma^{(6)}_{6}$, $\gamma^{(7)}_{13}=\gamma^{(6)}_{12}$, $\gamma^{(8)}_{4}=\gamma\{(\sqrt{2}\bar c(\bar c\bar d - 2\bar a\bar b)
+\bar a(\bar a\bar d+ \bar b\bar c)/\sqrt{2})(1+\chi_{r})+\frac{\bar d}{2\sqrt{2}}(1-\chi_{r})\}$, $\gamma^{(8)}_{8}=\gamma\{(4(\bar a\bar b-\bar c\bar d)^{2} 
+ 2(\bar a\bar d + \bar b\bar c)^{2} + \bar a^{2}/2 + \bar c^{2}/2)(1+\chi_{r})+(\bar d^{2}+\bar b^{2})(1-\chi_{r})/2\}$,
$\gamma^{(9)}_{5}=\gamma^{(8)}_{4}$, $\gamma^{(9)}_{9}=\gamma^{(8)}_{8}$, $\gamma^{(10)}_{10}
=\gamma\{(4(\bar c\bar d)^{2} + (\bar a\bar d + \bar b\bar c)^{2}+\bar a^{2}/2)(1+\chi_{r})+(1/2+\bar d^{2})(1-\chi_{r})/2\}$,
$\gamma^{(11)}_{0}=2\gamma\{(3\sqrt{2}\bar d\bar c^{2}+\bar a(\bar a\bar d+ \bar b\bar c)/\sqrt{2})(1+\chi_{r})+\frac{\bar d}{2\sqrt{2}}(1-\chi_{r})\}$,
$\gamma^{(11)}_{1}=2\sqrt{2}\gamma \bar d \bar c^{2}(1+\chi_{r})$, $\gamma^{(11)}_{2}=2\gamma\{(3\sqrt{2}\bar d\bar c^{2}+\bar a(\bar a\bar d 
+ \bar b\bar c)/\sqrt{2})(1+\chi_{r})-\frac{\bar d}{2\sqrt{2}}(1-\chi_{r})\}$, $\gamma^{(11)}_{3}=2\gamma\{(3\sqrt{2}\bar d\bar c^{2} - \bar a(\bar a\bar d+
 \bar b\bar c)/\sqrt{2})(1+\chi_{r})+ \frac{\bar d}{2\sqrt{2}}(1-\chi_{r})\}$,
$\gamma^{(11)}_{11}=\gamma\{(4(\bar c\bar d)^{2} + (\bar a\bar d + \bar b\bar c)^{2}+\bar a^{2}/2 +2\bar c^{2})(1+\chi_{r})+(1/2+\bar d^{2})(1-\chi_{r})/2\}$,
$\gamma^{(12)}_{6} = \gamma^{(13)}_{7} = \gamma^{(8)}_{4}$, $\gamma^{(12)}_{12}=\gamma^{(13)}_{13}=\gamma^{(8)}_{8}$.

\end{document}